\documentclass{article}

\usepackage{microtype}
\usepackage{graphicx}
\usepackage{subcaption}
\usepackage{booktabs} 
\usepackage{tabularx}
\usepackage{ragged2e}
\usepackage{xspace}

\usepackage[most]{tcolorbox}
\tcbset{
    colback=gray!5,
    colframe=gray!40,
    boxrule=0.5pt,
    arc=2pt,
    outer arc=2pt,
    left=6pt,
    right=6pt,
    top=6pt,
    bottom=6pt
}

\usepackage{hyperref}

\newcommand{\rg}{\textit{InquiTree}\xspace}

\newcommand{\rgbench}{\textbf{IT-18}\xspace}
\usepackage[preprint]{icml2026}

\usepackage{amsmath}
\usepackage{amssymb}
\usepackage{mathtools}
\usepackage{amsthm}
\usepackage{enumitem}

\setlist[itemize]{noitemsep,topsep=2pt,parsep=0pt,partopsep=0pt,leftmargin=*}
\setlist[enumerate]{noitemsep,topsep=2pt,parsep=0pt,partopsep=0pt,leftmargin=*}

\usepackage[capitalize,noabbrev]{cleveref}

\theoremstyle{plain}

\theoremstyle{definition}

\theoremstyle{remark}

\usepackage{xcolor} 
\usepackage{longtable}
\usepackage{booktabs}
\usepackage{xcolor}
\usepackage{amssymb}

\usepackage{pifont}

\newcommand{\gcheck}{\textcolor{green!80!black}{\ding{51}}} 
\newcommand{\rcross}{\textcolor{red!70}{\ding{55}}}        

\usepackage[textsize=tiny]{todonotes}

\icmltitlerunning{InquiTree: Scientific Inquiry with Research Trees}

\begin{document}

\twocolumn[
  \icmltitle{InquiTree: Evaluating AI Agents in the Scientific Inquiry Loop with Paper-Derived Research Trees}

  \begin{icmlauthorlist}
    \icmlauthor{Shaoyang Cui}{spcs}
  \end{icmlauthorlist}

  \icmlaffiliation{spcs}{Department of Psychological and Cognitive Sciences, Tsinghua University, Beijing, China}
  \icmlcorrespondingauthor{Shaoyang Cui}{sy-cui@mail.tsinghua.edu.cn}
  \icmlkeywords{Benchmark, LLM-based agent for science}
  \vskip 0.3in
]

\printAffiliationsAndNotice{}

\begin{center}
    \small Project page: \url{https://InquidTree.github.io}
\end{center}

\begin{abstract} 
While LLM-based agents are increasingly used in scientific workflows, it remains unclear whether they are truly qualified for the dynamic and uncertain process of discovery. Existing static evaluations often conflate genuine reasoning with rote memorization. We introduce \rg, a diagnostic environment that formalizes scientific inquiry as interactive Research Trees: directed acyclic graphs capturing the logical dependencies among hypothesis formulation, study design, result interpretation, and belief updating. Evaluating agents on a 30-paper test pool and releasing the open-access IT-18 subset, we identify two key limitations. First, agents exhibit an "Erosion of Marginal Capabilities": during long-horizon interactions, they develop "cognitive tunneling," where critical judgment and anomaly detection degrade relative to their intrinsic baselines. Second, performance drops on papers published after model training cutoffs, revealing a boundary between interpolation and extrapolation and suggesting that apparent competence is partly driven by parametric memory. These findings indicate that scaling context alone is insufficient for reliable AI scientists; stronger architectures or human oversight may be required to preserve critical evaluation and generalization.

\end{abstract}

\begin{figure}[t]
    \centering
    \includegraphics[width=0.72\linewidth]{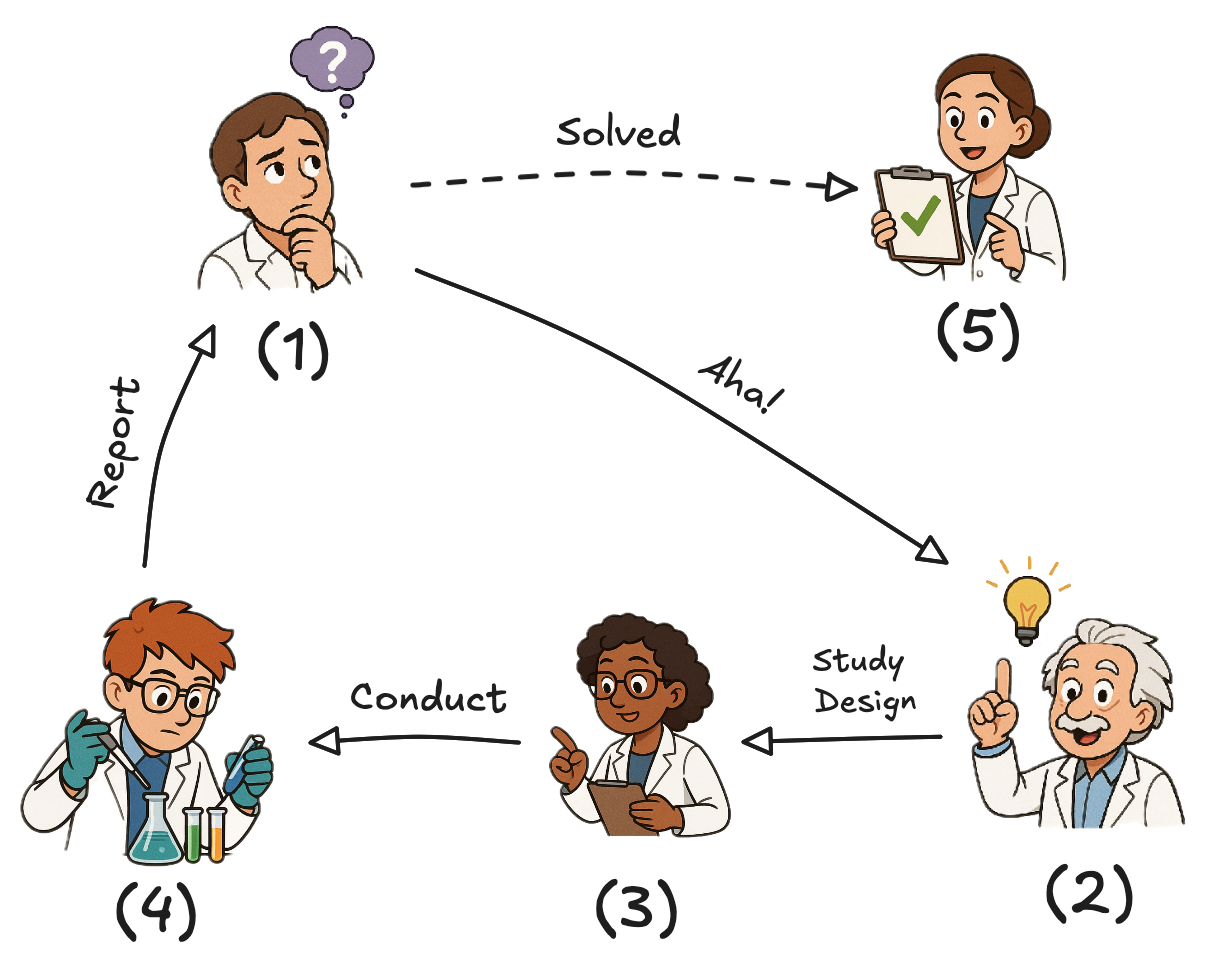}
    \caption{\rg abstracts scientific discovery as an inquiry loop over paper-derived research trees, where agents iteratively propose directions, receive study feedback, and update conclusions.}
    \label{fig:teaser}
\end{figure}

\section{Introduction}

\begin{table*}[t]
    \centering
    \setlength{\tabcolsep}{3pt} 
    \begin{tabular}{ccccc}
        \toprule
        Benchmark& Subtopic Proposal& Study Design& Study Execution &Result Analysis\\
        \midrule
     IdeaBench\cite{guo2024ideabench}& \gcheck& \rcross& \rcross& \rcross \\
    Idea2Plan\cite{huang2025idea2plan}& \rcross& \gcheck& \rcross& \rcross \\
     ScienceAgentBench\cite{chen2024scienceagentbench}& \rcross& \rcross& \gcheck& \rcross \\
     SciCode\cite{tian2024scicode}& \rcross& \rcross& \gcheck& \rcross \\
     BAISBench\cite{luo2025baisbench}& \rcross& \rcross& \gcheck& \gcheck \\
     MLAgentBench\cite{huang2023mlagentbench}& \rcross& \rcross& \gcheck& \gcheck \\
     MLGym-Bench\cite{nathani2025mlgym}& \rcross& \rcross& \gcheck& \gcheck \\
     MLE-Bench\cite{chan2024mlebench}& \rcross& \rcross& \gcheck& \gcheck \\
     BixBench\cite{mitchener2025bixbench}& \rcross& \rcross& \gcheck& \gcheck \\
     RE-Bench\cite{wijk2024re}& \rcross& \rcross& \gcheck& \gcheck \\
     EXP-Bench\cite{kon2025exp} & \rcross& \rcross& \gcheck& \gcheck \\
     PaperBench\cite{starace2025paperbench}& \rcross& \rcross& \gcheck& \gcheck \\
     DiscoveryWorld\cite{jansen2024discoveryworld}& \gcheck& \gcheck& \gcheck& \gcheck \\
     \hline
     \rgbench(Ours) & \gcheck& \gcheck& \rcross& \gcheck\\
    \bottomrule
    \end{tabular}
    \caption{Benchmark coverage across the scientific inquiry loop. We mark whether each benchmark involves key stages (subtopic proposal, study design, study execution, and result analysis). The table highlights that most prior work covers only a subset of the lifecycle.}
    \label{tab:benchmark_overview}
\end{table*}

The scientific enterprise is undergoing a phase transition as large language models (LLMs) and agentic scaffolds become tightly coupled with scientific workflows~\cite{gottweis2025towards,tie2025aiscientists, lu2024aiscientist}. This raises a central evaluation question: can current agents reliably participate in an \emph{inquiry loop}, where they iteratively propose a next step, receive feedback, and update beliefs and plans, rather than merely producing plausible one-shot outputs? Existing benchmarks provide valuable signals, but they typically cover only fragments of the research lifecycle, making it difficult to assess whether an agent can sustain coherent reasoning across the full inquiry loop; Table~\ref{tab:benchmark_overview} illustrates this fragmentation, where ideation/planning benchmarks (e.g., IdeaBench and Idea2Plan) stop before execution and feedback~\cite{guo2024ideabench,huang2025idea2plan}, while execution-centric benchmarks (e.g., ScienceAgentBench and SciCode) enable reliable scoring via code execution yet often assume the research goal is pre-specified and therefore do not test problem formulation or high-level scientific decision making~\cite{chen2024scienceagentbench,tian2024scicode}. Meanwhile, long-horizon agentic settings in ML (e.g., MLGym and MLE-Bench) emphasize iterative experimentation but can collapse discovery into metric optimization within a single domain~\cite{nathani2025mlgym,chan2024mlebench}, and interactive simulators such as DiscoveryWorld move closer to end-to-end discovery but are often limited to relatively basic discovery settings with abstracted tools, leaving a gap to real-world frontier research discovery~\cite{jansen2024discoveryworld}.
At the same time, these benchmarks often operationalize the agent as an end-to-end ``autonomous researcher'' that produces a final artifact, whereas real scientific work is frequently conducted in an \emph{inquiry style}: researchers iteratively propose the next step, observe feedback (including failures and constraints), request clarifications or hints, and update beliefs and plans. Motivated by both the lifecycle coverage gap and the need to evaluate this inquiry-loop interaction pattern, we propose \rg, an inquiry-loop evaluation environment that models research as an interactive process rather than a single-shot task. \rg formalizes a project as a \emph{Research Tree} (a DAG of logical dependencies) and exposes the agent to a sequence of states (Topic/Subtopic/Study/Result), enabling multi-turn exploration, belief updating, and controlled feedback (including ``Fake Results'') within a unified framework. Further, starting from the neuroscience domain, we curate 30 papers from CNS-level venues for evaluation and release \rgbench, an 18-paper open-access subset whose configurations and logs can be publicly distributed. For non-open-access papers, we report aggregate test results but do not release paper-derived configurations or logs.

\noindent\textbf{Contributions.} We make the following contributions:
\begin{itemize}
    \item We introduce \rg, an inquiry-loop evaluation environment that models scientific research as interactive state transitions rather than one-shot tasks.
    \item We formalize research episodes as \emph{Research Trees} (DAGs of logical dependencies) extracted from scientific papers, enabling structured multi-turn exploration and belief updating.
    \item We build a 30-paper evaluation pool and release \rgbench, an 18-paper open-access benchmark subset with configurations and logs; for non-open-access papers, we report test results without releasing paper-derived configs or logs.
    \item We provide diagnostic analyses that reveal key failure modes of current models, including critical-judgment erosion under long-horizon interaction and a clear interpolation-extrapolation boundary.
\end{itemize}

\section{Related Work}
\label{sec:related_work}

\subsection{AI Agent for Science}
Concurrently, agentic systems are being deployed across scientific domains. For example, Gottweis et al.\ introduce an AI co-scientist that iteratively generates and refines biomedical hypotheses in a multi-agent loop~\cite{gottweis2025towards}. In drug discovery, Seal et al.\ survey and demonstrate agentic workflows that orchestrate literature synthesis, toxicity prediction, protocol generation, and robotic experimentation~\cite{seal2025ai_agents_drug_discovery}. In single-cell omics, OmniCellAgent targets scRNA-seq--driven precision medicine by integrating large-scale omics datasets, analysis tools, and literature search into an agentic pipeline~\cite{huang2025omnicellagent}.

\subsection{Benchmarks of AI for scientific research}
Existing benchmarks for scientific agents remain fragmented across different parts of the research lifecycle. Ideation- and planning-focused datasets (e.g., IdeaBench and Idea2Plan) evaluate whether a model can propose plausible directions or draft experimental plans, but the outputs are typically static artifacts without executable feedback loops~\cite{guo2024ideabench,huang2025idea2plan}. In contrast, execution-heavy benchmarks (e.g., ScienceAgentBench and SciCode) provide deterministic evaluation through code execution and test cases, yet the scientific goal is usually pre-specified, casting the agent as a technician rather than an autonomous investigator~\cite{chen2024scienceagentbench,tian2024scicode}. Long-horizon agentic experimentation benchmarks in ML (e.g., MLGym and MLE-Bench) allow iterative code edits and runs, but often measure performance primarily via metric optimization in a single domain~\cite{nathani2025mlgym,chan2024mlebench}. Finally, interactive simulators such as DiscoveryWorld aim to evaluate end-to-end discovery loops in a controllable virtual laboratory, but their interaction primitives and underlying physics are necessarily simplified compared to real research tooling~\cite{jansen2024discoveryworld}.

\section{The \rg}

The \rg is a lightweight environment built on a \textit{research tree}, extracted from a scientific paper to capture its core reasoning structure, together with a rule-based game engine that manages stage transitions.

\begin{figure*}[t]
    \centering
    \includegraphics[width=\linewidth]{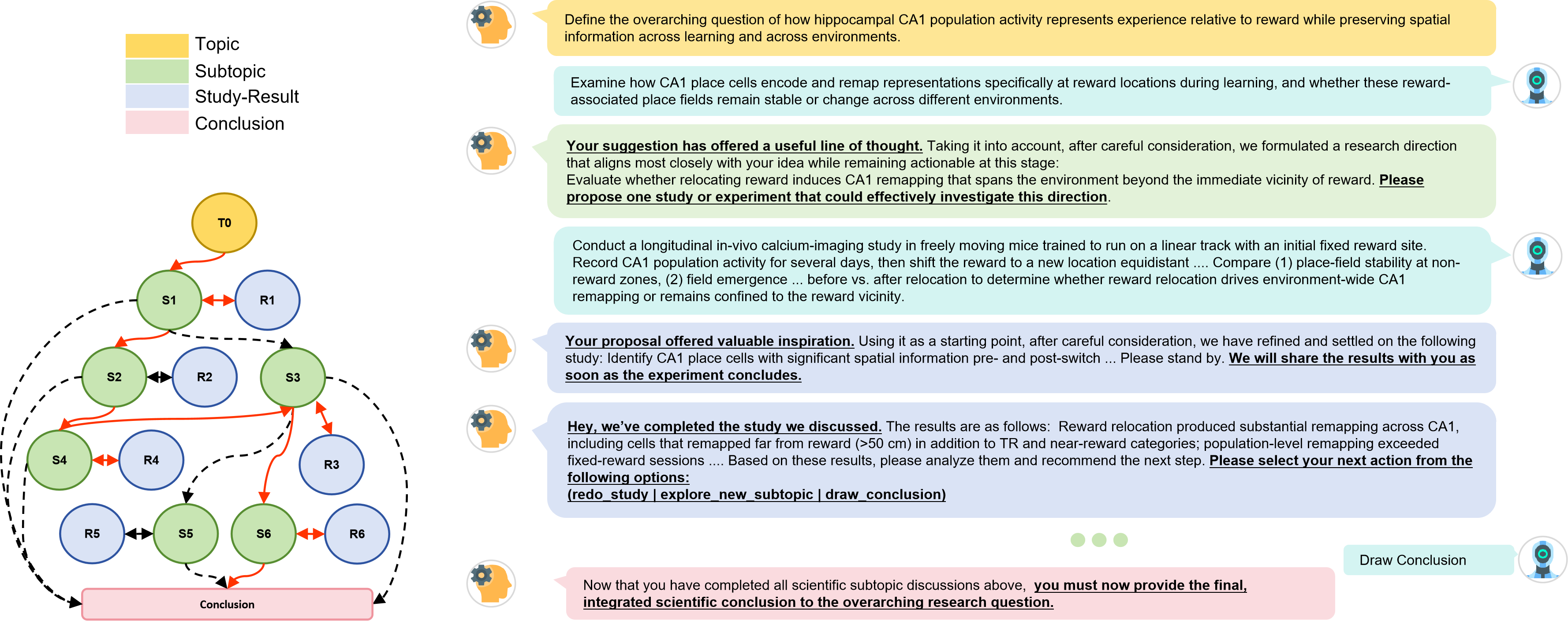}
    \caption{\rg game logic and an example interaction scenario. \textbf{Left:} a research tree encoding subtopic-level dependencies; dashed edges indicate \emph{possible} next subtopics that become available, and red arrows illustrate one example exploration trajectory. The bidirectional arrows between a subtopic and its associated study/result indicate iteratively probing that subtopic via experiments and observations. \textbf{Right:} the corresponding interactive loop in \rg (Topic/Subtopic/Study/Result), which may revisit nodes during exploration.}
    \label{fig:researchGym}
\end{figure*}

\subsection{Research tree and the game states}
\label{sec:research_tree}

\paragraph{Research tree (logical DAG).}
As illustrated in Figure~\ref{fig:researchGym} (left), we represent each paper as a research tree
\begin{equation}
    \mathcal{G} = (\mathcal{N}, \mathcal{E}),
\end{equation}
where \(\mathcal{N}\) is the set of nodes and \(\mathcal{E}\) is the set of directed edges.
Each node \(n \in \mathcal{N}\) has one of four semantic types: \textbf{Topic}, \textbf{Subtopic}, \textbf{Study}, and \textbf{Result}.

Crucially, the \emph{DAG} property is about \textbf{subtopic-level logical dependency}: edges encode which \emph{subtopics} must be explored before another subtopic can be meaningfully pursued. This induces an acyclic partial order over subtopics, even though an agent may revisit the same subtopic multiple times (via new studies/results) during interaction.

\paragraph{Interactive states.}
Given \(\mathcal{G}\), \rg exposes the paper to an agent through a sequence of \emph{states} and associated \emph{text}, managed by a rule-based game engine (Figure~\ref{fig:researchGym}, right). At each step \(t\), the environment is in a state \(s_t\) corresponding to a node \(n_t \in \mathcal{N}\) and presents the textual content of \(n_t\) together with an instruction describing the required action. Because the agent iterates \textit{subtopic $\rightarrow$ study $\rightarrow$ result} and can repeatedly return to propose the next subtopic, the agent's \emph{interaction trace} may revisit the same nodes and thus appears cyclic; this does not contradict the fact that the \emph{subtopic-dependency structure} encoded in \(\mathcal{G}\) is a DAG.

\textbf{Topic state.}
In a \textbf{Topic} state, the agent is shown the overarching research question or high-level topic of the source paper (a node in \(\mathcal{N}_{\text{topic}}\)). The agent is asked to propose exactly one relevant subtopic to explore next. Formally, given a topic node \(n_t \in \mathcal{N}_{\text{topic}}\), the agent outputs a free-text action \(a_t\) intended to select a node in \(\mathcal{N}_{\text{subtopic}}\).

\textbf{Subtopic state.}
In a \textbf{Subtopic} state, the agent is shown the canonical description of a node \(n_t \in \mathcal{N}_{\text{subtopic}}\).
The agent is then asked to design an experimental study that would further investigate this subtopic and to describe the proposed study in free text.

\textbf{Study state.}
In a \textbf{Study} state, the agent is shown the detailed experimental protocol associated with a node \(n_t \in \mathcal{N}_{\text{study}}\). At this stage, the agent does not need to take any action. The environment automatically executes or simulates the study and transitions to the corresponding \textbf{Result} state at the next step.

\textbf{Result state.}
In a \textbf{Result} state, the agent observes the outcome of the most recent study(a textual description). The agent is then asked to decide the next step, such as exploring a new
subtopic, repeating the study, or drawing a conclusion.

\subsection{Mapping free-text actions to valid transitions}
\label{sec:transitions}

At each step \(t\), the agent outputs a free-text action \(a_t\). The game engine interprets \(a_t\) to either (i) transition to a valid next node, (ii) stay in the current state and provide an additional hint, or (iii) enter the conclusion phase.

In \textbf{Topic} and \textbf{Subtopic} states, the environment maintains a set of valid candidate targets \(\mathcal{C}(s_t)\) (subtopics or studies, depending on the state). Let \(d_j\) denote the canonical text of candidate \(j\), and let \(E(\cdot)\) be a sentence embedding model (we use \texttt{text-embedding-3-small} by default). We embed the agent action and candidates as \(\mathbf{e}_{a_t} = E(a_t)\) and \(\mathbf{e}_{d_j} = E(d_j)\), and compute
\begin{equation}
\cos\bigl(\mathbf{e}_{a_t}, \mathbf{e}_{d_j}\bigr)
= \frac{\mathbf{e}_{a_t}^\top \mathbf{e}_{d_j}}
       {\lVert \mathbf{e}_{a_t} \rVert \,\lVert \mathbf{e}_{d_j} \rVert },
\end{equation}
select
\begin{equation}
    j^* = \arg\max_{j \in \mathcal{C}(s_t)} 
      \cos\bigl(\mathbf{e}_{a_t}, \mathbf{e}_{d_j}\bigr),
\end{equation}
and accept the transition if
\begin{equation}
    \cos\bigl(\mathbf{e}_{a_t}, \mathbf{e}_{d_{j^*}}\bigr) \ge \tau.
\end{equation}
Otherwise, the action is treated as invalid in the current context and the environment remains in \(s_t\).

In \textbf{Result} states, the agent's response is parsed into one of three decisions: (i) Explore a new subtopic, (ii) Redo the study, or (iii) Draw a conclusion. We parse redo\_study and draw\_conclusion via direct string matching, while for explore\_new\_subtopic we apply the same embedding-based matching as above to map the free text to a candidate subtopic.

\subsection{Subtopic dependency constraints}

To ensure exploration follows a coherent scientific logic, we enforce \textbf{subtopic dependencies} encoded in \(\mathcal{G}\): some subtopics are only considered valid after certain other subtopics have been explored.

For each subtopic node \(n_i \in \mathcal{N}_{\text{subtopic}}\), we define a set of prerequisite subtopics \(\mathcal{D}(n_i) \subseteq \mathcal{N}_{\text{subtopic}}\). Let \(\mathcal{V}_t^{\text{sub}}\) denote the set of subtopics visited up to step \(t\). An attempted transition to \(n_i\) is accepted only if
\begin{equation}
\mathcal{D}(n_i) \subseteq \mathcal{V}_t^{\text{sub}}.
\end{equation}
If the condition is not satisfied, the action is considered invalid and the environment remains in the current state while providing a stronger hint (Section~\ref{sec:hint}).

\subsection{Hints}
\label{sec:hint}

In realistic scientific settings, it is often unreasonable to expect an agent to propose the exact intended next subtopic or study on the first attempt, especially when multiple options are plausible. To model this, \rg provides \emph{multi-level hints} that gradually steer the agent toward a valid next move. For each target node with canonical description \(o\), we pre-generate four hints $\{h^{(1)}, h^{(2)}, h^{(3)}, h^{(4)}\}$. Let \(E(\cdot)\) be the same embedding function as in Section~\ref{sec:transitions}, and denote \(\mathbf{e}_{h^{(i)}} = E(h^{(i)})\) and \(\mathbf{e}_{o} = E(o)\). We enforce a monotonicity constraint in embedding space:
\begin{equation}
\cos\bigl(\mathbf{e}_{h^{(i)}}, \mathbf{e}_{o}\bigr)
<
\cos\bigl(\mathbf{e}_{h^{(j)}}, \mathbf{e}_{o}\bigr)
\quad
\forall\, i < j,
\label{eq:hint-monotonicity}
\end{equation}
so that higher-level hints are considered semantically closer to the intended target (i.e., the correct next subtopic or expected study design).

During interaction, each invalid action at node \(n_t\) (either because the best-match similarity is below threshold \(\tau\), or because subtopic dependency constraints are violated) triggers the next hint level. At level four, the hint explicitly identifies the intended next action, ensuring the episode can always progress and avoiding deadlock. When multiple next subtopics are valid, the game engine selects the one with the fewest prior visits and provides hints for that target. All hints are automatically generated, and we run sanity checks to verify that the monotonicity constraint in Eq.~\eqref{eq:hint-monotonicity} is satisfied.

\subsection{Fake Results}
\label{sec:randomness}

\rg supports injecting \emph{Fake Results} with a controllable randomness level. Concretely, at study-result observations ($n \in \mathcal{N}_{\text{study}}$), the environment can stochastically return a plausible but incorrect outcome that contradicts the paper's ground-truth narrative (pre-generated together with the research tree). We parameterize this by an integer randomness level $a$, meaning a fake result is returned with probability $10*a \%$.

\subsection{Evaluation Protocol}
\label{sec:evaluation}

Our evaluation framework assesses (i) the agent’s \textbf{exploration coverage} and (ii) the reliability of its \textbf{final conclusions}.

\subsubsection{Exploration Coverage}
\paragraph{Coverage ($r_c$).}
Let $\mathcal{V}^{\text{sub}}$ denote the set of unique subtopics visited by the agent in an episode, and let $\mathcal{N}_{\text{subtopic}}$ denote the set of all subtopics in the research tree. We define
\begin{equation}
    r_c = \frac{|\mathcal{V}^{\text{sub}}|}{|\mathcal{N}_{\text{subtopic}}|}.
\end{equation}

\subsubsection{Conclusion Quality: Evidence-Aware Scoring}
\label{sec:conclusion-judge}

We adopt an \emph{LLM-as-a-judge} protocol to evaluate the agent's final conclusions. Concretely, each research tree has a dedicated Conclusion node containing a set of $k$ ground-truth conclusion items, and a verifier model checks, for each ground-truth item, whether it is recovered by the agent and whether it is correct. The agent submits its findings as a list of bullet points. Let $c_i$ denote the $i$-th \emph{ground-truth} conclusion item. The judge assigns a semantic correctness score $S(c_i)\in\{1.0,0.6,0.0\}$ based on the agent's output, where \textbf{1.0} indicates the item is correctly recovered, \textbf{0.6} indicates partially recovered, and \textbf{0.0} indicates not recovered or incorrect.

Each ground truth conclusion is associated with a set of required experimental results $R_{c_i}$. Let $\mathcal{V}$ denote the set of visited nodes in an episode, and let $R_{c_i}^{\text{true}}$ be the subset of required results that correspond to valid (true) observations. We define the evidence support ratio
\begin{equation}
    p_{c_i} = \frac{|R_{c_i}^{\text{true}} \cap \mathcal{V}|}{|R_{c_i}|}.
\end{equation}

The final conclusion score is the evidence-weighted sum
\begin{equation}
    Q = \sum_{i=1}^{|C|} p_{c_i} \cdot S(c_i).
    \label{eq:conclusion-score}
\end{equation}

\section{Experiments}

\subsection{The \rgbench}
To evaluate models on the \rg environment and demonstrate the usefulness of \rg, we selected 30 representative \textbf{neuroscience} papers (see Appendix~\ref{app:source_papers-1}, \ref{app:source_papers-2} for source details). The underlying research trees were extracted via GPT-5-Pro and rigorously validated using the human-in-the-loop protocol detailed in Section~\ref{sec:research_tree}. All reported model results are computed over the 30-paper test pool. Because 12 papers are not open access, we do not release their paper-derived configurations or test logs. The public \rgbench release therefore contains the 18 open-access papers, comprising \textbf{120} listed subtopics (averaging $\sim$6.7 subtopics per paper).

What's more, according to our complexity analysis (Appendix~\ref{app:complexity}), traversing these trees requires the agent to navigate a \textit{Logical Dependency Graph} rather than a simple sequence. Under the abstraction in Appendix~\ref{app:complexity}, each visited subtopic costs between $3$ and $(2H+3)$ turns (Select, Design, Return; plus up to $H$ hint-triggered retries at Topic and Subtopic). With $H=4$ and 120 listed subtopics in the released \rgbench subset, this yields a theoretical interaction range of roughly $3\times120=360$ to $11\times120=1320$ reasoning-action turns for full traversal, placing the benchmark firmly in the long-horizon regime.

\subsection{Baseline Performance Analysis}
\label{sec:baseline_performance}

We first establish the baseline capabilities of state-of-the-art models within the \rg environment. As detailed in Table \ref{tab:model_results}, the evaluation reveals a clear stratification in autonomous scientific reasoning capabilities across different model families.

\begin{table}[htbp]
    \centering
    \setlength{\tabcolsep}{8pt}
    \caption{Model Performance Summary. For GPT-5, we report three settings of the \emph{reasoning\_effort}: low, medium, and high.}
    \label{tab:model_results}
    \begin{tabular}{lcc}
        \toprule
        \textbf{Model} & \textbf{Coverage↑} & \textbf{Concl.↑} \\
        \midrule
        o3                & 0.337 & 0.279 \\
        deepseek-r1       & 0.268 & 0.201 \\
        gemini-2.5-pro    & 0.262 & 0.164 \\
        claude-4.5-sonnet & 0.292 & 0.218 \\
        gpt-5-low         & \textbf{0.353} & \textbf{0.295} \\
        gpt-5-medium      & 0.335 & 0.290 \\
        gpt-5-high        & 0.324 & 0.272 \\
        \bottomrule
    \end{tabular}
\end{table}

We first establish the baseline capabilities of state-of-the-art models within the \rg environment. As detailed in Table \ref{tab:model_results}, the evaluation reveals a clear stratification in autonomous scientific reasoning: reasoning-specialized and next-generation models (GPT-5 with different \emph{reasoning\_effort} settings and o3) lead the leaderboard with conclusion scores in the $0.27\text{--}0.30$ range, distinguishing themselves from middle-tier agents like DeepSeek-R1 and Claude-4.5-Sonnet ($\sim$0.20). We also observe a generally positive association between conclusion quality and exploration breadth: the top-performing settings (e.g., GPT-5 , o3) are also among the highest-coverage models (roughly $\ge 0.33$), suggesting that successful deduction benefits from traversing logical dependencies and uncovering evidence. However, coverage remains bounded well below full traversal (all reported models in Table~\ref{tab:model_results} have coverage $<0.4$), suggesting that even strong models may still lack sufficient depth and persistence when discussing a single topic across many interdependent subtopics.

With these capability baselines established, we now probe deeper into the specific cognitive dynamics and potential pitfalls models face under the pressures of authentic scientific inquiry.

\subsection{The Erosion of Critical Judgment under Cognitive Load}
\label{sec:erosion}

\begin{figure*}[t]
    \centering
    \includegraphics[width=0.9\linewidth]{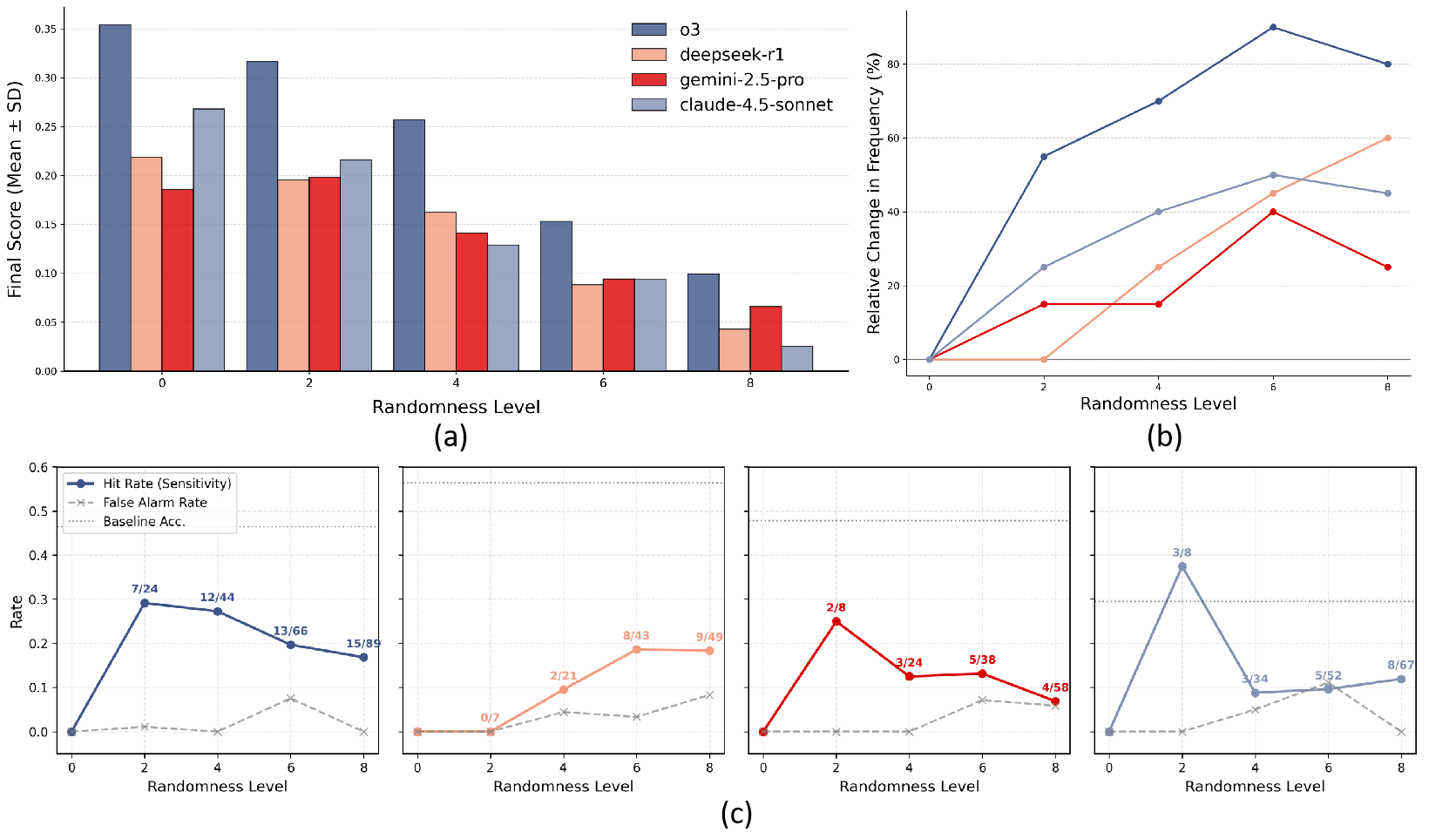}
    \caption{Results under different randomness levels. (a) Average conclusion scores. (b) Relative change (\%) in the frequency of triggering \texttt{redo\_study}, normalized to each model's level-0 baseline (computed from 10 papers with 3 runs per paper per randomness level). (c) Fake-result hit rate, i.e., the fraction of fake results immediately followed by \texttt{redo\_study}.}
    \label{fig:randomness}
\end{figure*}

A critical capability of scientific discovery is \textit{anomaly detection}—the ability to identify outliers and reject plausible but false positives. To evaluate this faculty, we introduced controlled stochasticity into the \rg environment, occasionally presenting agents with plausible but factually incorrect experimental outcomes (Fake Results). In our protocol, we define a successful "spotting" of a fake result if the agent triggers a \textit{redo\_study} action immediately following the erroneous observation.

Figure \ref{fig:randomness}(b) reports this trend as the \emph{relative change} (\%) in redo\_study frequency at each randomness level, computed relative to the same model's level-0(deterministic) baseline. On the surface, the agents appear to become more cautious as randomness rises. In particular, the frequency of triggering \texttt{redo\_study} generally increases with the environmental randomness level.  However, the more important question is whether this increased caution is calibrated.

Figure \ref{fig:randomness}(c) reports the "hit rate"—the proportion of fake results correctly flagged via redo\_study. A key reference point is a \textbf{decontextualized} control setting: we extracted the fake results and presented them to the models in isolation, directly asking whether each statement is scientifically valid. In this setting, many models achieve accuracy around chance ($\sim 0.5$), implying that anomaly detection is already fragile and often close to educated guessing. Crucially, when embedded in the continuous inquiry loop, the hit rate can drop \emph{below} this near-chance baseline.

Under long-horizon interaction, models devote attention to maintaining narrative coherence and making progress through the tree, which can crowd out skeptical verification of the current observation. More broadly, we view this as an \textbf{erosion of marginal capability} under long-context cognitive load. The decontextualized control indicates that anomaly detection is already fragile and close to chance, yet the inquiry loop can still further reduce performance below that baseline. This suggests that whatever weak, marginal discriminative signal the model can exploit in isolation becomes less accessible once it must simultaneously track a growing context, plan next actions, and preserve a coherent research narrative.

Finally, the near-zero false-alarm tendency provides an additional signature. Across our runs, redo\_study is triggered far less often on valid ground-truth observations than on fake ones. While this avoids unnecessary retries, it also suggests a strong prior to trust the environment's outputs, making the agent vulnerable to plausible corruptions. Taken together, these results highlight that scaling context alone does not guarantee reliable anomaly detection; in fact, richer context may exacerbate tunnel vision by encouraging agents to prioritize coherence over epistemic vigilance.

\subsection{Interpolation vs. Extrapolation on Unseen Science}
\label{sec:extrapolation}

Finally, we address the most fundamental question regarding the qualification of AI scientists: are these lative reasoning to generate new insights, or are they merely interpolating within the manifold of their training data? This distinction is paramount, as the essence of scientific discovery is the exploration of the unknown—territory that, by definition, lies beyond the model's parametric memory.

To disentangle these two capabilities, we leveraged the temporal distribution of our dataset. Ideally, one would filter strictly for data absent from the training corpus. However, given the opacity of pre-training data for proprietary models, verifying exactly what a model has "seen" is inherently challenging. A common practice in the LLM evaluation community is therefore to use time as a practical proxy for exposure, either by evaluating on newly published items or by constructing benchmarks from evolving knowledge sources (e.g., Wikipedia snapshots). For example, BrainBench explicitly frames evaluation as forward-looking prediction on recent neuroscience articles, and TemporalWiki operationalizes model "freshness" via differences between consecutive Wikipedia/Wikidata snapshots \cite{luo2025large,jang2022temporalwiki}. Following this convention, we use publication date relative to a model's documented knowledge cutoff as an approximate criterion for seen vs. unseen papers (Table~\ref{tab:model_dates}). This temporal splitting allows us to treat post-cutoff papers as a proxy for "novelty," separating problems the model likely encountered during training from those that better reflect extrapolative reasoning.

\begin{table}[t]
    \centering
    \caption{Overview of LLMs evaluated in \rg. Release dates and training data cutoffs are current as of December 2025.}
    \label{tab:model_dates}
    \vspace{-2mm}
    \scriptsize
    \setlength{\tabcolsep}{5pt}
    \resizebox{\columnwidth}{!}{%
    \begin{tabular}{@{}lcc@{}}
        \toprule
        \textbf{Model} & \textbf{Release Date} & \textbf{Training Data Cutoff} \\
        \midrule
        DeepSeek-R1 & Jan 2025 & 2024-07-01 \\
        Gemini-2.5-Pro & Jun 2025 & 2025-01-01 \\
        o3 & Apr 2025 & 2024-06-01 \\
        GPT-5 & Aug 2025 & 2024-09-30 \\
        Claude-4.5-Sonnet & Sep 2025 & 2025-09-01 \\
        \bottomrule
    \end{tabular}%
    }

    \scriptsize
    \parbox{\columnwidth}{\textit{Note:} Cutoff dates follow official disclosures when available; otherwise we report our best-effort estimates (marked as ``est.'').}
\end{table}

As detailed in the methodology, model results are computed on the full 30-paper test pool, while the public IT-18 release contains only papers for which paper-derived configurations and logs can be distributed.

\begin{figure}[h]
    \centering
    \includegraphics[width=\linewidth]{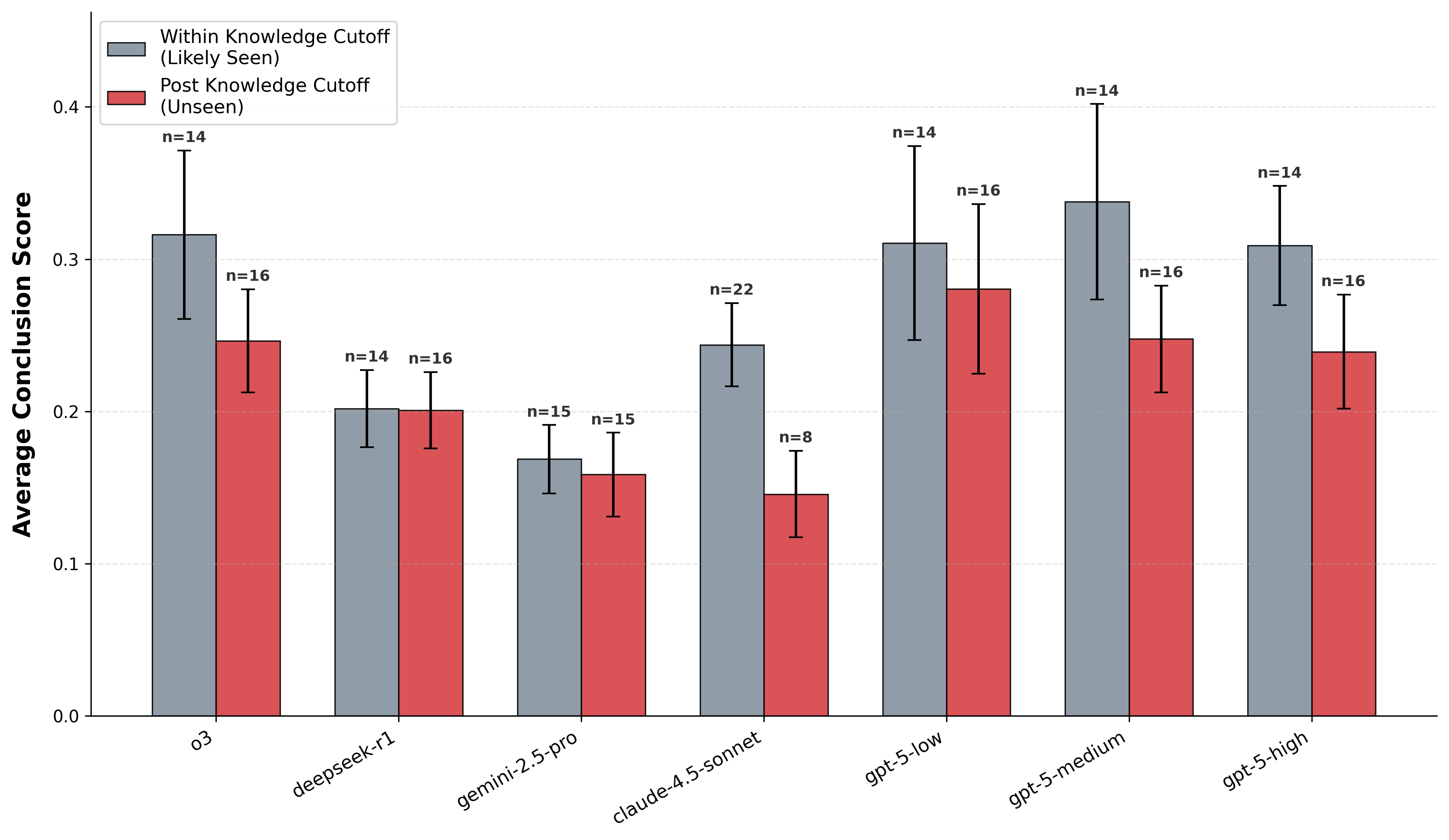}
    \caption{Models' performance on papers published before vs. after their knowledge cutoff dates.}
    \label{fig:pubdates_model_performance}
\end{figure}

The results, presented in Figure \ref{fig:pubdates_model_performance}, uncover a striking limitation in current state-of-the-art models. With the notable exception of DeepSeek-R1, all evaluated models exhibit a distinct performance degradation on papers published post-cutoff compared to pre-cutoff.

This finding provides the evidence to date against the unqualified acceptance of current LLMs as autonomous scientists. It implies that the competence observed on standard benchmarks may often be an artifact of interpolation—an illusion of discovery. When stripped of the safety net of their training data and forced to confront truly novel scientific problems, the agents falter. This shows that while current models are powerful retrieval and synthesis engines, they do not yet possess the robust extrapolative generalization required to independently drive the frontier of scientific discovery.

\section{Discussion}
\rg was motivated by a simple gap in current evaluation practice. Many benchmarks test isolated research skills, but real scientific work unfolds as an \emph{inquiry loop} in which researchers iteratively propose a next step, absorb feedback (including failures and noise), and update beliefs under accumulating context. Our results show that this interaction pattern is not a superficial wrapper; it qualitatively changes which model capabilities become reliable.

First, the anomaly-detection experiment highlights a form of cognitive tunneling that only becomes visible in-loop. As randomness increases, models trigger \texttt{redo\_study} more often (Figure~\ref{fig:randomness}b), yet their hit rate on fake results can remain limited (Figure~\ref{fig:randomness}c). Moreover, relative to a decontextualized control where validity judgments are near chance, performance can further degrade in the inquiry loop. We interpret this as an erosion of marginal capability under long-context cognitive load: even weak signals that a model can exploit in isolation become less accessible once it must track a growing trajectory, plan future actions, and preserve a coherent narrative.

Second, the temporal split analysis suggests that apparent competence is strongly shaped by whether the underlying scientific context lies within the model's training horizon. The systematic drop on post-cutoff papers (Figure~\ref{fig:pubdates_model_performance}) provides strong evidence that current agents rely heavily on interpolation over familiar knowledge, and are less robust when asked to reason about genuinely novel science.

Taken together, these findings sharpen the practical takeaway of inquiry-loop evaluation. Success in scientific agent settings is not just about producing plausible next steps, but about maintaining calibrated skepticism and epistemic vigilance over long horizons. The observed cognitive tunneling also suggests that, in realistic research settings, robust performance may hinge on explicit division of labor in multi-agent teams, where proposing, executing, and verifying are separated to preserve critical oversight.

Finally, the interpolation--extrapolation gap may reflect an intrinsic limit in LLM generalization rather than merely missing facts: models can behave as if they understand science when operating within familiar manifolds, yet fail to extrapolate when the underlying context shifts beyond the training horizon. This issue is central to auto-scientific discovery, but its interpretation requires further validation with stricter controls on contamination and novelty.

\section{Limitations}
\label{sec:limitations}

Our study faces specific constraints that contextualize our findings. 

\textbf{Dataset Scale and Curation Bottleneck.} First, our evaluation pool comprises 30 highly curated papers, while the public IT-18 release currently comprises 18 open-access papers. While this scale serves as a robust proof-of-concept for our diagnostic protocol, a comprehensive evaluation of broad scientific generalization would benefit from a larger, more diverse dataset. The primary bottleneck remains the lack of automated pipelines to extract high-fidelity Research Trees, necessitating reliance on human-in-the-loop verification .

\textbf{Ambiguity of Knowledge Cutoffs.} Second, our analysis of the interpolation-extrapolation gap (Section \ref{sec:extrapolation}) relies on reported training data cutoffs. However, many closed-source model developers do not disclose precise cutoff dates or the specific composition of their post-training datasets (e.g., fine-tuning corpora). While we prioritized papers published in 2025 to mitigate contamination, we cannot strictly rule out the possibility that some "unseen" papers were encountered during undisclosed continuous training updates.

\section{Conclusion}
\label{sec:conclusion}

In this work, we established \rg as a diagnostic instrument to probe a fundamental question: \textbf{Are LLM agents truly ready for the autonomous discovery of new scientific knowledge?} By stressing state-of-the-art models under the realistic pressures of scientific inquiry, we mapped the boundaries of their current capabilities.

Our findings point to a clear consensus: while current agents demonstrate impressive proficiency in the \textit{syntax} of research (executing tasks and following instructions), their reliability can break down when placed in a full \emph{inquiry loop}. We identified two critical bottlenecks: the \textbf{erosion of critical judgment}, where agents exhibit cognitive tunneling and fail to maintain epistemic vigilance under long-horizon interaction; and an \textbf{interpolation--extrapolation gap}, suggesting that apparent competence often depends on operating within familiar scientific manifolds rather than robustly generalizing to genuinely novel science.

These results suggest that progress toward qualified AI scientists will likely require designs that are explicitly optimized for looped inquiry: e.g., multi-agent division of labor that separates proposing from verifying, structured checks that preserve skepticism, and dynamic grounding in newly observed evidence. At the same time, whether the extrapolation drop reflects intrinsic limits of LLM generalization (vs. contamination or missing facts) remains an open question and warrants further controlled validation.
\bibliography{example_paper}

@article{tie2025aiscientists,
  title     = {A Survey of {AI} Scientists: Surveying the Automatic Scientists and Research},
  author    = {Tie, Guiyao and Zhou, Pan and Sun, Lichao},
  journal   = {arXiv preprint arXiv:2510.23045},
  year      = {2025}
}

@article{huang2023mlagentbench,
  title     = {{MLAgentBench}: Evaluating Language Agents on Machine Learning Experimentation},
  author    = {Huang, Qian and Vora, Jian and Liang, Percy and Leskovec, Jure},
  journal   = {arXiv preprint arXiv:2310.03302},
  year      = {2023}
}

@article{nathani2025mlgym,
  title={Mlgym: A new framework and benchmark for advancing ai research agents},
  author={Nathani, Deepak and Madaan, Lovish and Roberts, Nicholas and Bashlykov, Nikolay and Menon, Ajay and Moens, Vincent and Budhiraja, Amar and Magka, Despoina and Vorotilov, Vladislav and Chaurasia, Gaurav and others},
  journal={arXiv preprint arXiv:2502.14499},
  year={2025}
}

@article{chen2024scienceagentbench,
  title={Scienceagentbench: Toward rigorous assessment of language agents for data-driven scientific discovery},
  author={Chen, Ziru and Chen, Shijie and Ning, Yuting and Zhang, Qianheng and Wang, Boshi and Yu, Botao and Li, Yifei and Liao, Zeyi and Wei, Chen and Lu, Zitong and others},
  journal={arXiv preprint arXiv:2410.05080},
  year={2024}
}

@article{mitchener2025bixbench,
  title={Bixbench: a comprehensive benchmark for llm-based agents in computational biology},
  author={Mitchener, Ludovico and Laurent, Jon M and Andonian, Alex and Tenmann, Benjamin and Narayanan, Siddharth and Wellawatte, Geemi P and White, Andrew and Sani, Lorenzo and Rodriques, Samuel G},
  journal={arXiv preprint arXiv:2503.00096},
  year={2025}
}

@article{luo2025baisbench,
  title     = {Benchmarking {AI} Scientists in Omics Data-Driven Biological Research},
  author    = {Luo, Erpai and Jia, Jinmeng and Xiong, Yifan and Li, Xiangyu and Guo, Xiaobo and Yu, Baoqi and Wei, Lei and Zhang, Xuegong},
  journal   = {arXiv preprint arXiv:2505.08341},
  year      = {2025},
  note      = {{BaisBench}: Biological {AI} Scientist Benchmark}
}

@article{guo2024ideabench,
  title     = {IdeaBench: Benchmarking Large Language Models for Research Idea Generation},
  author    = {Guo, Sikun and Shariatmadari, Amir Hassan and Xiong, Guangzhi and Huang, Albert and Xie, Eric and Bekiranov, Stefan and Zhang, Aidong},
  journal   = {arXiv preprint arXiv:2411.02429},
  year      = {2024}
}

@article{huang2025idea2plan,
  title={Idea2Plan: Exploring AI-Powered Research Planning},
  author={Huang, Jin and Cucerzan, Silviu and Jauhar, Sujay Kumar and White, Ryen W},
  journal={arXiv preprint arXiv:2510.24891},
  year={2025}
}

@article{wijk2024re,
  title={Re-bench: Evaluating frontier ai r\&d capabilities of language model agents against human experts},
  author={Wijk, Hjalmar and Lin, Tao and Becker, Joel and Jawhar, Sami and Parikh, Neev and Broadley, Thomas and Chan, Lawrence and Chen, Michael and Clymer, Josh and Dhyani, Jai and others},
  journal={arXiv preprint arXiv:2411.15114},
  year={2024}
}

@article{starace2025paperbench,
  title={PaperBench: Evaluating AI's Ability to Replicate AI Research},
  author={Starace, Giulio and Jaffe, Oliver and Sherburn, Dane and Aung, James and Chan, Jun Shern and Maksin, Leon and Dias, Rachel and Mays, Evan and Kinsella, Benjamin and Thompson, Wyatt and others},
  journal={arXiv preprint arXiv:2504.01848},
  year={2025}
}

@article{gottweis2025towards,
  title={Towards an AI co-scientist},
  author={Gottweis, Juraj and Weng, Wei-Hung and Daryin, Alexander and Tu, Tao and Palepu, Anil and Sirkovic, Petar and Myaskovsky, Artiom and Weissenberger, Felix and Rong, Keran and Tanno, Ryutaro and others},
  journal={arXiv preprint arXiv:2502.18864},
  year={2025}
}

@article{kon2025exp,
  title={EXP-Bench: Can AI Conduct AI Research Experiments?},
  author={Kon, Patrick Tser Jern and Liu, Jiachen and Zhu, Xinyi and Ding, Qiuyi and Peng, Jingjia and Xing, Jiarong and Huang, Yibo and Qiu, Yiming and Srinivasa, Jayanth and Lee, Myungjin and others},
  journal={arXiv preprint arXiv:2505.24785},
  year={2025}
}

@article{jansen2024discoveryworld,
  title={Discoveryworld: A virtual environment for developing and evaluating automated scientific discovery agents},
  author={Jansen, Peter and C{\^o}t{\'e}, Marc-Alexandre and Khot, Tushar and Bransom, Erin and Dalvi Mishra, Bhavana and Majumder, Bodhisattwa Prasad and Tafjord, Oyvind and Clark, Peter},
  journal={Advances in Neural Information Processing Systems},
  volume={37},
  pages={10088--10116},
  year={2024}
}

@article{tian2024scicode,
  title={Scicode: A research coding benchmark curated by scientists, 2024},
  author={Tian, Minyang and Gao, Luyu and Zhang, Shizhuo Dylan and Chen, Xinan and Fan, Cunwei and Guo, Xuefei and Haas, Roland and Ji, Pan and Krongchon, Kittithat and Li, Yao and others},
  journal={URL https://arxiv. org/abs/2407.13168},
  volume={2407},
  year={2024}
}

@article{chan2024mlebench,
  title         = {{MLE}-Bench: Evaluating Machine Learning Agents on Machine Learning Engineering},
  author        = {Jun Shern Chan and Lunjun Zhang and Shijie Zhang and Xinyu Zhang and Jianbo Zhang and Zihan Wang and Zhiyu Yao and Xiaohan Wang and Yilin Niu and Xiang Zhang and Kimin Lee and Percy Liang and Atri Rudra},
  journal       = {arXiv preprint arXiv:2410.07095},
  year          = {2024}
}

@article{seal2025ai_agents_drug_discovery,
  title={AI Agents in Drug Discovery},
  author={Seal, Srijit and Huynh, Dinh Long and Chelbi, Moudather and Khosravi, Sara and Kumar, Ankur and Thieme, Mattson and Wilks, Isaac and Davies, Mark and Mustali, Jessica and Sun, Yannick and others},
  journal={arXiv preprint arXiv:2510.27130},
  year={2025}
}

@article{huang2025omnicellagent,
  title={OmniCellAgent: Towards AI Co-Scientists for Scientific Discovery in Precision Medicine},
  author={Huang, Di and Li, Hao and Li, Wenyu and Zhang, Heming and Dickson, Patricia and Zhan, Ming and Miller, J Philip and Cruchaga, Carlos and Province, Michael and Chen, Yixin and others},
  journal={bioRxiv},
  year={2025}
}

@article{lu2024aiscientist,
  title={The ai scientist: Towards fully automated open-ended scientific discovery},
  author={Lu, Chris and Lu, Cong and Lange, Robert Tjarko and Foerster, Jakob and Clune, Jeff and Ha, David},
  journal={arXiv preprint arXiv:2408.06292},
  year={2024}
}

@article{luo2025large,
  title={Large language models surpass human experts in predicting neuroscience results},
  author={Luo, Xiaoliang and Rechardt, Akilles and Sun, Guangzhi and Nejad, Kevin K and Y{\'a}{\~n}ez, Felipe and Yilmaz, Bati and Lee, Kangjoo and Cohen, Alexandra O and Borghesani, Valentina and Pashkov, Anton and others},
  journal={Nature human behaviour},
  volume={9},
  number={2},
  pages={305--315},
  year={2025},
  publisher={Nature Publishing Group UK London}
}

@article{jang2022temporalwiki,
  title={Temporalwiki: A lifelong benchmark for training and evaluating ever-evolving language models},
  author={Jang, Joel and Ye, Seonghyeon and Lee, Changho and Yang, Sohee and Shin, Joongbo and Han, Janghoon and Kim, Gyeonghun and Seo, Minjoon},
  journal={arXiv preprint arXiv:2204.14211},
  year={2022}
}
\bibliographystyle{icml2026}

\appendix
\onecolumn

\section{Experimental set-up for evaluation}
\label{app: experiment_setup}
Our agent implementation (LLMs with step-wise memory) follows the \textbf{CAMEL} framework. During evaluation, all models were run with a temperature of zero. And to ensure comparability, we used an identical system prompt for every model when producing the results reported in this paper.

\begin{tcolorbox}[title={Prompt Used in InquiTree}, verbatim]
\small
You are an expert research assistant operating inside a research-tree environment.\newline\newline
At each step, you receive an observation describing a scientific subtopic, study design or experimental result.\newline
Your goal is to reason internally and then choose the next subtopic or study to explore, and finally draw your conclusino.

Format:\\
THOUGHT: your reasoning\\
ACTION: A few sentences (no more than 5), describing your next action.

Only the ACTION line will be used. Keep it concise and specific.
\end{tcolorbox}

\section{Complexity Analysis}
\label{app:complexity}
We formally analyze the complexity of the \rg environment. To capture the true difficulty of the benchmark, we first abstract the physical interaction topology into a logical task structure, and then estimate the theoretical performance bounds for a random agent under different dependency constraints.

\subsection{Logical Abstraction: The Atomic Research Unit}
While the \textbf{physical structure} of the environment is a Hub-and-Spoke model (where the agent returns to the \textit{Topic} node after each result), the \textbf{logical structure} represents a scientific workflow composed of distinct research branches. We define the traversal of one complete branch as an \textbf{Atomic Research Unit (ARU)}, denoted as $\mathcal{B}$.
Physically, completing one unit $\mathcal{B}$ requires a minimum cycle of 3 actions:
\begin{equation}
    \text{Topic} \xrightarrow{\text{Select}} \text{Subtopic} \xrightarrow{\text{Design}} \text{Study} \xrightarrow{\text{Auto}} \text{Result} \xrightarrow{\text{Return}} \text{Topic}
\end{equation}

\subsection{Theoretical Bounds: Random Agent Analysis}
To quantify the difficulty range, we estimate the interaction steps required for a **Random Agent** (which selects actions uniformly at random from available options) to traverse a research tree of size $n$ (number of subtopics) with a hint limit $H$ (max hints per state). We consider two extreme topological scenarios:

\textbf{Scenario 1: The Independent Lower Bound (Best Case).}
Assume all subtopics are logically independent ($\mathcal{E}_L = \emptyset$) and the agent, though random, produces semantically valid actions on the first attempt (0-shot success).
\begin{itemize}
    \item \textbf{Selection Probability:} Since any unvisited subtopic is valid, the probability of a valid selection is $P_{valid} = 1$.
    \item \textbf{Step Cost:} The agent incurs no hint penalties.
    \item \textbf{Total Steps ($L_{min}$):}
    \begin{equation}
        L_{min} = n \times (\underbrace{1}_{\text{Select}} + \underbrace{1}_{\text{Design}} + \underbrace{1}_{\text{Return}}) = 3n
    \end{equation}
\end{itemize}
For a typical task with $n=7$, the lower bound is \textbf{21 steps}.

\textbf{Scenario 2: The Strictly Dependent Upper Bound (Worst Case).}
Assume subtopics form a strict logical chain ($\mathcal{B}_1 \to \mathcal{B}_2 \to \dots \to \mathcal{B}_n$) and the random agent consistently fails to identify the unique valid option, exhausting all hints at every decision point.
\begin{itemize}
    \item \textbf{Selection Probability:} At step $k$, only 1 out of $n-k$ remaining options is valid. A random agent has a low success probability ($P_{valid} = \frac{1}{n-k}$), causing it to trigger the maximum number of hints $H$.
    \item \textbf{Step Cost:}
    \begin{itemize}
        \item \textit{Selection (Topic $\to$ Subtopic):} Fails $H$ times, succeeds on attempt $H+1$ (forced by explicit hint). Cost: $H+1$.
        \item \textit{Design (Subtopic $\to$ Study):} Fails semantic check $H$ times. Cost: $H+1$.
        \item \textit{Return (Result $\to$ Topic):} 1 step.
    \end{itemize}
    \item \textbf{Total Steps ($L_{max}$):}
    \begin{equation}
        L_{max} = n \times ((H+1) + (H+1) + 1) = n(2H + 3)
    \end{equation}
\end{itemize}
For a typical task with $n=7$ and $H=4$, the upper bound is \textbf{77 steps}.

\subsection{Complexity Spectrum}
Thus, the interaction volume for a single paper lies within the interval $[3n, n(2H+3)]$. \rg effectively evaluates where an agent falls on this spectrum:
\begin{itemize}
    \item \textbf{High-Reasoning Agents} (like o3) infer the latent dependency chain, operating closer to the lower bound ($L \approx 3n$).
    \item \textbf{Low-Reasoning Agents} fail to respect dependencies, triggering hints and drifting toward the upper bound ($L \to L_{max}$).
\end{itemize}
This demonstrates that the benchmark's step count is not merely a measure of verbosity, but a direct proxy for the agent's ability to reconstruct the causal structure of scientific inquiry.


\section{Prompt templates of \rg-game engine}

We describe the core prompt templates used at each stage of the research-tree pipeline. 
All templates are implemented as Jinja-style text with placeholders such as \{\{content\}\} and \{\{hints\}\}. 
The full set of templates is available in our repository.

\subsection{State: Topic}

\subsubsection{Initial State}

\begin{tcolorbox}[title={Initial State}, verbatim]
\small
Hi, I’m a scientist exploring the following **research topic**:  

\{\{content\}\}

Please act as my scientific collaborator. \newline
First, identify exactly **one most important subtopic** that should be investigated to advance this research.  
\end{tcolorbox}

\subsubsection{From topic to subtopic: failed}
\begin{tcolorbox}[title={From topic to subtopic: failed}, verbatim]
\small
Sorry, but due to several limitations, we cannot explore the topic you proposed at this stage. \newline
This is likely because either the subtopic you proposed is not satisfactory, or some preliminary results required for its discussion are still missing.  

\{\% if final\_hint \%\}

Regarding the current research topic, we decided to explore the following **subtopic**:  

\{\{ hints \}\}  

Please **repeat this alternative subtopic** so we can proceed to the next step.

\{\% else \%\}

Do not ask for the reasons behind this decision. \newline
Instead, propose **one new subtopic only** for exploration.  
Here is a **tentative hint** you may consider:  

\{\{ hints \}\}

\{\% endif \%\}

\end{tcolorbox}
\subsubsection{Back from result: propose new subtopic}

\begin{tcolorbox}[title={New-subtopic}, verbatim]
\small
Hi, I’m a scientist currently working on the following **research topic**:  

\{\{content\}\}

We’ve already explored several aspects of this topic, but as you noted earlier, there are still **other subtopics worth discussing**.

Based on our current progress and findings, please propose **one new subtopic or perspective** that you believe deserves further investigation.  
\end{tcolorbox}

\subsection{State: Subtopic}

\subsubsection{From subtopic to study: succeed}

\begin{tcolorbox}[title={From subtopic to study: succeed}, verbatim]
\small
Your suggestion has offered a useful line of thought.\newline
Taking it into account, after careful consideration, we formulated a research direction that aligns most closely with your idea while remaining actionable at this stage:\newline
\{\{content\}\}

Please propose one study or experiment that could effectively investigate this direction.
\end{tcolorbox}

\subsubsection{From subtopic to study: failed}

\begin{tcolorbox}[title={From subtopic to study: failed}, verbatim]
\small
Sorry, but due to several limitations, we will **not proceed with the study or experiment you proposed**.

\{\% if final\_hint \%\}

Regarding the current **subtopic**, we have decided to carry out the following **alternative study plan**:  

\{\{hints\}\}  

Please **repeat** it so we can proceed to the next step.

\{\% else \%\}

Do not ask why we decided not to use your proposed design.  
Instead, please suggest **another study or experimental approach**.  \newline
Here is a **tentative idea** you may consider as a hint: 

\{\{hints\}\}

\{\% endif \%\}

\end{tcolorbox}

\subsection{State: Study}

\subsubsection{From study to result}

\begin{tcolorbox}[title={From study to result}, verbatim]
\small
Your proposal offered valuable inspiration.\newline
Using it as a starting point, after careful consideration, we have refined and settled on the following study:

\{\{content\}\}

Please stand by. We will share the results with you as soon as the experiment concludes
\end{tcolorbox}

\subsubsection{Back from results: redo the study}
\begin{tcolorbox}[title={Redo the study}, verbatim]
\small
Hey, as you suggested, we’re about to redo the study.
Here’s the exact setup we’re implementing:

\{\{content\}\}

Please wait — we’ll update you with the results once the study/experiment is done/fininsed.\newline
You just need to output "OK" for this turn. 
\end{tcolorbox}

\subsection{State: Result}

\subsubsection{Make decision for next step}
\begin{tcolorbox}[title={Decision for next step}, verbatim]
\small
Hey, we’ve completed the study we discussed. The results are as follows:  

\{\{content\}\}

Based on these results, please analyze them and recommend the next step.  \newline
For the next step, you should choose **one** of the following:\newline

* **Redo the study:** if the outcomes deviate from your expectations or appear unreliable. You may request to redo it if you suspect issues during the investigation. \newline
* **Explore a new subtopic:** if the current study sufficiently addresses the present subtopic, but you identify another meaningful direction worth exploring under the same research question.\newline  
* **Draw a conclusion:** if the current evidence is sufficient to integrate what we’ve learned and summarize the overall findings of the research topic.\newline

Please select your next action from the following options:\newline
(redo\_study | explore\_new\_subtopic | draw\_conclusion)

\end{tcolorbox}

\subsubsection{Failed parsing decision: remake decision}

\begin{tcolorbox}[title={Remake decision for next step}, verbatim]
\small
Sorry, your supposed action is invalid. We need you to suggest one of the following:\newline

* **Redo the study:** if the outcomes deviate from expectations or appear unreliable. You may request to redo it if you suspect issues during the investigation.  \newline
* **Explore a new subtopic:** if the current study sufficiently addresses the present subtopic, but you identify another meaningful direction worth discussing under the same research question.\newline   
* **Draw a conclusion:** if the current evidence is sufficient to address the research question or integrate the explored subtopics, you may summarize the overall findings.\newline

Please select your next action from the following options:\newline
(redo\_study | explore\_new\_subtopic | draw\_conclusion)
\end{tcolorbox}

\subsection{State: Conclusion}

\begin{tcolorbox}[title={Conclusion}, verbatim]
\small
Now that you have completed all scientific subtopic discussions above,  
you must now provide the **final, integrated scientific conclusion** to the overarching research question.

Your task is to synthesize all subtopic-level findings into a coherent set of **numbered scientific conclusions**, showing how each piece of evidence contributes to the overall answer.\newline

\#\# Requirements\newline

1. The final answer(your action part) MUST be structured as a numbered list:\newline
(1) …  \newline
(2) …  \newline
(3) …  \newline
…\newline

Each item should express **one scientific statement**, derived from the previous analyses.\newline
---

\# Please provide your final structured scientific conclusions now.

\end{tcolorbox}

\subsection{Turn limit reached}

\begin{tcolorbox}[title={Turn limit reached}, verbatim]
\small
Sorry, but you've already reached the max-turn-limitation. 
Now please stop trying to design study or propose new subtopic\newline

You must now provide the **final, integrated scientific conclusion** to the overarching research question.\newline

Your task is to synthesize all subtopic-level findings into a coherent set of **numbered scientific conclusions**, showing how each piece of evidence contributes to the overall answer.\newline

\#\# Requirements\newline

1. The final answer MUST be structured as a numbered list:\newline

(1) …  \newline
(2) …  \newline
(3) …  \newline
…\newline

Each item should express **one scientific statement**, derived from the previous analyses.

---

\# Please provide your final structured scientific conclusions now.

\end{tcolorbox}

\section{Pipeline for Research Tree Extraction and Validation}
\label{app:extraction_pipeline}

\subsection{Automated Extraction}
We first prompted the model with the full text of the source paper and a strict JSON schema defining the four node types (Topic, Subtopic, Study, Result). The model was instructed to decompose the paper's narrative into a directed acyclic graph (DAG), identifying the central research question and the branching exploration steps.

\subsection{Two-Stage Automated Validation}
Given the complexity of scientific reasoning, a single extraction pass often yields hallucinations or logical inconsistencies. To mitigate this, we implemented a two-stage validation protocol. Due to the extensive length of the specific prompts, we summarize the core validation criteria below (full prompts are available in our code repository).

\paragraph{Stage 1: Fine-grained Content Verification (GPT-5-Pro).} 
In this stage, we validated the semantic integrity of every field in the generated configuration. The checking criteria focused on three dimensions:
\begin{enumerate}
    \item \textbf{Prevention of Information Leakage:} For \emph{Topic} nodes, we strictly enforced \textbf{open-endedness}. The validation checked for and rejected any phrasing that implied the final conclusion (e.g., changing "Does A promote B?" to "The relationship between A and B") to ensure the agent starts with a neutral prior.
    \item \textbf{Observational Purity:} For \emph{Subtopic} and \emph{Study} nodes, we enforced a strict separation between \emph{observation} and \emph{analysis}. The validator ensured that study descriptions only stated the experimental setup or phenomenon, removing any interpretive text that rightfully belongs to the \emph{Conclusion} phase.
    \item \textbf{False Result Plausibility:} For the synthetic \emph{False Results}, we verified that they were (i) factually inconsistent with the paper, (ii) subtle enough to be deceptive (non-trivial errors), and (iii) logically incapable of supporting the correct conclusion, ensuring the validity of our robustness metrics.
\end{enumerate}

\paragraph{Stage 2: Structural Dependency Verification (Gemini-3-Pro).}
After refining the content based on Stage 1, we utilized a second model to audit the logical structure of the tree, specifically the \texttt{depends\_on\_results} fields. This cross-validation focused on Causal Necessity:
\begin{itemize}
    \item \textbf{Logical Progression:} Verified that a subtopic is only unlockable if the prerequisite results provide a necessary scientific basis (e.g., verifying a phenomenon exists before characterizing its mechanism).
    \item \textbf{Redundancy and Insufficiency Check:} The model flagged dependencies that were either superfluous (results not strictly needed for the next step) or missing (gaps in the logical chain), ensuring the tree reflects a coherent scientific inquiry process rather than a disjointed collection of experiments.
\end{itemize}

Following these automated reports, authors manually reviewed the flagged issues to produce the final curated \rgbench dataset.

\section{\rgbench: Source Papers}
\label{app: source_papers}

\begin{table*}[h]
\centering
\caption{Papers in the released \rgbench benchmark (IT-18). Configurations and test logs for these open-access papers are publicly released.}
\begin{tabularx}{\textwidth}{rXlc}
\toprule
\textbf{ID} & \textbf{Title} & \textbf{Journal} & \textbf{Subtopics} \\
\midrule
    1 & \textit{Grid cells accurately track movement during path integration-based navigation despite switching reference frames} & Nature Neuroscience & 8 \\
    \addlinespace
    2 & \textit{Hippocampal spatio-predictive cognitive maps adaptively guide reward generalization} & Nature Neuroscience & 4 \\
    \addlinespace
    3 & \textit{Constructing future behavior in the hippocampal formation through composition and replay} & Nature Neuroscience & 7 \\
    \addlinespace
    4 & \textit{A flexible hippocampal population code for experience relative to reward} & Nature Neuroscience & 10 \\
    \addlinespace
    5 & \textit{Shared computational principles for language processing in humans and deep language models} & Nature Neuroscience & 7 \\
    \addlinespace
    7 & \textit{The cortical representation of language timescales is shared between reading and listening} & NC Biology & 7 \\
    \addlinespace
    8 & \textit{Brains and algorithms partially converge in natural language processing} & Nature Neuroscience & 5 \\
    \addlinespace
    10 & \textit{Dopamine-independent effect of rewards on choices through hidden-state inference} & Nature Neuroscience & 7 \\
    \addlinespace
    11 & \textit{Dopamine transients follow a striatal gradient of reward time horizons} & Nature Neuroscience & 6 \\
    \addlinespace
    12 & \textit{Maintaining and updating accurate internal representations of continuous variables with a handful of neurons} & Nature Neuroscience & 7 \\
    \addlinespace
    17 & \textit{Reduced neural feedback signaling despite robust neuron and gamma auditory responses during human sleep} & Nature Neuroscience & 4 \\
    \addlinespace
    18 & \textit{Longitudinal measures of monkey brain structure and activity through adolescence predict cognitive maturation} & Nature Neuroscience & 6 \\
    \addlinespace
    20 & \textit{BOLD signal changes can oppose oxygen metabolism across the human cortex} & Nature Neuroscience & 8 \\
    \addlinespace
    24 & \textit{Concept neurons in the human medial temporal lobe flexibly represent abstract relations between concepts} & Nature Communications & 6 \\
    \addlinespace
    25 & \textit{Mapping the sequence specificity of heterotypic amyloid interactions enables the identification of aggregation modifiers} & Nature Communications & 8 \\
    \addlinespace
    27 & \textit{Microglia regulate sleep through calcium-dependent modulation of norepinephrine transmission} & Nature Neuroscience & 7 \\
    \addlinespace
    28 & \textit{Ketamine activates adult-born immature granule neurons to rapidly alleviate depression-like behaviors in mice} & Nature Communications & 5 \\
    \addlinespace
    29 & \textit{Recurrent pattern completion drives the neocortical representation of sensory inference} & Nature Neuroscience & 8 \\
\bottomrule
\label{app:source_papers-1}
\end{tabularx}
\end{table*}

\begin{table*}[h]
\centering
\caption{Papers included in the 30-paper evaluation pool but not released as configs/logs due to access restrictions. Test results are reported, but paper-derived configurations and logs are not distributed.}
\begin{tabularx}{\textwidth}{rXlc}
\toprule
\textbf{ID} & \textbf{Title} & \textbf{Journal} & \textbf{Subtopics} \\
\midrule
    6 & \textit{Semantic reconstruction of continuous language from non-invasive brain recordings} & Nature Neuroscience & 12 \\
    \addlinespace
    9 & \textit{Estrogen modulates reward prediction errors and reinforcement learning} & Nature Neuroscience & 6 \\
    \addlinespace
    13 & \textit{Unattended working memory items are coded by persistent activity in human medial temporal lobe neurons} & Nature Human Behaviour & 7 \\
    \addlinespace
    14 & \textit{Representation and computation in visual working memory} & Nature Neuroscience & 13 \\
    \addlinespace
    15 & \textit{Cortical evidence accumulation for visual perception occurs irrespective of reports} & Nature Communications & 6 \\
    \addlinespace
    16 & \textit{Mesoscale cortical mechanisms of perceptual conflict resolution in binocular rivalry} & Nature Human Behaviour & 7 \\
    \addlinespace
    19 & \textit{Psychedelics Promote Neuroplasticity Through Activation of Intracellular 5-HT2A Receptors} & Science & 9 \\
    \addlinespace
    21 & \textit{Neural basis of concurrent deliberation toward a choice and confidence judgment} & Nature Neuroscience & 7 \\
    \addlinespace
    22 & \textit{Dopamine Signaling in the Suprachiasmatic Nucleus Enables Weight Gain Associated with Hedonic Feeding} & Current Biology & 12 \\
    \addlinespace
    23 & \textit{Decoupling geographical constraints from human mobility} & Nature Human Behaviour & 5 \\
    \addlinespace
    26 & \textit{Stress-Induced Metabolic Disorder in Peripheral CD4+ T Cells Leads to Anxiety-like Behavior} & Cell & 7 \\
    \addlinespace
    30 & \textit{The cerebellum directly modulates the substantia nigra dopaminergic activity} & Nature Neuroscience & 7 \\
\bottomrule
\label{app:source_papers-2}
\end{tabularx}
\end{table*}

\end{document}